\documentclass[prc,aps,showpacs,floatfix,nofootinbib,twocolumn,letterpaper]{revtex4} 
\usepackage{graphicx} 
\usepackage{amssymb} 
\usepackage{amsmath} 
\usepackage{amsfonts} 

\usepackage{color}

\usepackage{ulem}

\def\be{\begin{equation}} 
\def\ee{\end{equation}}

\def\bfN{\boldsymbol{N}}
\def\bfH{\boldsymbol{H}}
\def\bfK{\boldsymbol{K}}
\def\bfGam{\boldsymbol{\Gamma}}
\def\bfV{\boldsymbol{V}}
\def\bfG{\boldsymbol{G}}

\begin{document} 

\title{Generator coordinate method for transition-state dynamics in nuclear
fission  }

\author{G.F. Bertsch}
\affiliation{ 
Department of Physics and Institute for Nuclear Theory, Box 351560, 
University of Washington, Seattle, Washington 98915, USA}

\author{K. Hagino}
\affiliation{ 
Department of Physics, Kyoto University, Kyoto 606-8502,  Japan} 

\begin{abstract}
Since its beginnings, fission theory has asumed that low-energy 
induced fission takes place through transition-state channels
at the barrier tops.  Neverthess,  up to now there is no microscopic
theory applicable to those conditions.  We suggest that modern
reaction theory is suitable for this purpose, and propose
a methodology based on a configuration-interaction framework
using the Generator Coordinate Method (GCM). Simple reaction-theoretic
models are constructed with the Gaussian Overlap 
Approximation (GOA) to 
parameterize both the dynamics
within the channels  and their incoherent couplings to 
states outside the barrier.  The physical characteristics of the
channels
examined here are their effective bandwidths 
and the quality of the coupling
to compound-nucleus states as measured by the transmission factor $T$.  
We also investigate the spacing of GCM states with respect to their degree
of overlap.  We find that a rather coarse mesh provides an acceptable 
accuracy for estimating the bandwidths and transmission factors.  The
common numerical stability problem in using the GCM is
avoided due to the choice of meshes and the finite bandwidths of the channels.  
The bandwidths of the channels are largely controlled by the zero-point
energy with respect to the collective coordinate in the GCM configurations.
\end{abstract}
 
\maketitle 
 
\section{Introduction}

An important goal in the description of fission
reactions is to understand their excitation functions, that
is, the probability that the
reaction leads to fission as a function of the total energy.  Another important
goal is to understand
the properties of the daughter nuclei after a fission event.
There has been enormous
progress in recent years on understanding the characteristics of the final
state thanks to improved computational
tools in many-body quantum mechanics such as the
time-dependent Hartree-Fock-Bogoliubov approximation \cite{whitepaper,bulgac2016}.

The theory of the excitation function in reactions
with many possible outcomes has not seen comparable advances.  Fission
theory has
relied on the transition-state hypothesis\footnote{ 
The term ``channel"  describes its
role in reaction theory better than ``state", but we shall use both
designations for the models presented here.} 
since the original paper by Bohr and Wheeler in 1939 \cite{bo39} 
and continuing up to the present era 
\cite{bo13,capote2009,chadwick2006,chadwick2011,cap2011,
lu2016,schmidt1991}. Briefly, it 
is encapsulated in the formula for the decay
rate $\Gamma_{\rm BW}$
\be
\Gamma_{\rm BW} = \frac{1}{2 \pi \rho} \sum_\mu T_\mu
\label{BW}
\ee 
where $\mu$ labels channels, $\rho$ is the level density of the compound
nucleus, and $T_\mu$ is a transmission coefficient or conductance.  It is
also identical to the penetration factor in subbarrier conductance. It
satisfies the bounds 
\be
0 \leq T_\mu \leq 1.
\label{bound}
\ee  
Typically the energy dependence of $T$ is assumed to be the same as
that of a particle traversing a one-dimensional potentiala barrier,
but that is a pure guess absent a microscopic understanding of the 
many-body Hamiltonian dynamics.

It is clear that the present time-dependent formulations
are ill-suited to the task of describing the
barrier-crossing dynamics in heavy
nuclei.  We expect that a formulation using reaction theory might be
more successful.  In this paper, we examine how the transition-state
dynamics might be realized in a reaction theory based on a 
configuration-interaction treatment of the Hamiltonian.

In the theory of large nuclei, one starts with
the wave functions of self-consistent
mean-field theory, such as those given by the energy density functionals
of Skyrme, Gogny, or relativistic formulations \cite{bender03}.  Besides the self-consistent
solutions of the Hartree-Fock (HF) or Hartree-Fock-Bogoliubov (HFB) equations,  an
adequate basis of states for studying transport properties can be
constructed using the Generator Coordinate Method (GCM).  This requires 
the calculation of  mean-field configurations that are constrained by
one or more  single-particle fields.
The GCM has been used previously for modeling fission dynamics near the barrier top
\cite{go05,re16,ta17}.  In those works, the authors used GCM with 
two constraining fields and the Gaussian Overlap Approximation (GOA) to
map the Hamiltonian onto a two-dimensional Schr\"odinger equation. 
However, the steps needed to arrive at a Schr\"odinger equation ignore 
the statistical aspects of the decay and gives no hint of a connection to 
Eq. (1). Here we only need the matrix elements between GCM configurations
in our reaction-theoretic approach, avoiding the mapping onto a Sch\"odinger
equation.

In an earlier paper \cite{be21} we showed how one can derive
the transition-state formula in a highly simplified configuration-interaction approach.  
Here we shall use the same reaction theory formalism to calculate transmission
coefficients, but with a more realistic description of the channels.
An important advantage of the reaction theory  is that 
statistical aspects of the theory can be easily included in the formalism
\cite{ha21}.
 
A technical obstacle in the GCM approach is the nonorthogonality of the basis
configurations. As will be shown, its formal difficulties are avoided 
by making use of the many-body Green's function defined in Eq.
(\ref{Green}) below. A related problem is the danger of 
numerical instabilities  when overlaps between configurations are large. 
We will show that this problem does not arise because
one can use coarse bases without much loss of accuracy.

For investigating transition channels in fermionic systems, the general
characteristics can be derived independently of the details of the
constraining  field.  A configuration is labeled by the expectation value
of the field; we shall call the expectation value $q_k$ 
for a $k$-th configuration in a finite-dimensional basis.

Besides the internal properties of the channel, one needs specific
information about the coupling to the reservoirs of states on either
side of the channel.  The situation is very similar to the cables
in computer networks. The cable has a characteristic
impedance, and conductance between the connected devices depends on impedance matching.
An optimally matched coupling yields a
transition conductance  $T=1$.  Mismatches decrease it and makes it
dependent on the signal frequency. For the fission theory, 
one needs to understand in  
detail the interaction connecting compound-nuclear states to the
states in the channel.  That is beyond the scale of this paper;
we will treat these couplings schematically.  

\section{GCM methodology for transmission channels}

The usual procedure for applying the GCM to nuclear spectroscopy consists of the following
steps.

1)  Select  a set of configurations calculated in mean-field
theory and constrained by some physical one-body field such as 
the mass quadrupole moment $Q$.  The set of expectation values of the
field  $(q_1,q_2,...q_{N})$ defines an $N$-dimensional 
basis for the configuration space.  In more advanced approaches
the configurations are projected to restore broken symmetries.

2)  Calculate the  matrix $\bfN$ of overlaps between 
configurations 
and the matrix $\bfH$ of the Hamiltonian or the energy functional
that plays the role of the Hamiltonian in the mean-field theory. Here and 
below we use boldface symbols for matrices.
              
3) Solve the hermitian  eigenvalue problem\footnote{The equation is
put into Hermitian form through the standard variable transformation
$\psi' = N^{1/2} \psi$.} 
(i.e., the Hill-Wheeler equation)
\be
\boldsymbol{H} \psi = E \bfN  \psi
\label{Hv=ESv}
\ee
for energies $E$ and corresponding $N$-dimensional wave functions $\psi$.

4)  Check for convergence by varying the number of configurations $N$ 
in the calculation.  The effect on the properties in the low-energy
part of the spectrum should be small.

Steps 1) and 2) are the same for calculating reaction rates in the
GCM, but the remaining steps are completely different. Namely,
the new steps are:  

3'.) In a new step 3), the Hamiltonian is made complex by adding
imaginary terms $-\bfGam_j/2$ to it and calculate the 
Green's function. This replaces the matrix diagonalization in the old step 3).
Each $\bfGam_j$ is a matrix of decay rates to states $j$ outside of the model
space. It is a sum of rank-one matrices, each corresponding to an $S$-matrix
channel \cite{ka15}. See Eq. (8) below for a consistent implementation in
our modeling framework.  Time-dependent methods also make use of complex 
Hamiltonians to 
treat fluxes out of the model spaces, as for in Ref. \cite{re16}, but there
the $\bfGam_j$ can be parameterized as diagonal matrices.
There are two decay modes in the present fission study,
one corresponding to the set
of compound nucleus states and the other to states in the second well and
beyond.  We
label $a$ and $b$, respectively, in the equation below.
The required Green's function is
\be
\bfG(E) = (\bfH -i\bfGam_a/2 -i\bfGam_b/2 - \bfN E)^{-1}.
\label{Green}
\ee

5.)  The 
transmission factor $T_{ab}$ between reservoir $a$ and $b$ is given by
$S$-matrix expression.   
\be
  T_{ab} = \sum_{\mu\in a}\sum_{\nu \in b} |S_{\mu\nu}|^2. 
\ee
The $S$-matrix is usually written in terms of $\bfH$ and a set of reduced
decay amplitudes as in \cite[Eq. 14-19]{ka15}.  In our application the
phase information in $S_{\mu\nu}$ is not needed and we use the
Datta formula \cite{da95,al21} to calculate $T_{ab}$ 
in terms of  $\bfG$ and $\bfGam$,
\be
T_{ab} = \sum_{jklm} (\bfGam_{a})_{jk}\bfG_{kl}(\bfGam_{b})_{lm} \bfG^*_{mim} = {\rm Tr}
\,\left(\bfGam_{a}\bfG\bfGam_{b}\bfG^*\right).
\label{datta}
\ee
The resulting  $T_{ab}$ is a continuous real function of $E$; the individual
channels  
satisfy $ 0 \leq T_{\mu\nu} \leq 1$.  
As in the procedure for spectroscopic studies, one gains
confidence by varying the dimension of the configuration spaces.

\section{Decay widths}

What is left now are the tasks of constructing the matrices $\bfN$, $\bfH$ 
and $\bfG$.  The overlap matrix $\bfN$ is simply the determinant of the
orbital overlaps when the configurations are pure Slater
determinants.  We shall not go into the well-known difficulties \cite{la09} in defining
$\bfH$ when the energetics are based on an energy functional rather than
a Hamiltonian, and simply remark that the prescriptions for dealing with
an energy density functional are well established.  

A new issue
arises in defining the decay matrix. Our guiding principle is Fermi's
Golden Rule for the decay of a configuration $j$ into a set of
states $a$. This reads
\be
\Gamma_{a,j} = 2\pi \sum_{k\in a} |\langle k| v | j\rangle|^2\delta(E_k - E),
\ee
where the density-of-states function $\sum_{k\in a} \delta(E_k -E)$ smeared
out in some way for numerical computations. 
The state $j$ is in the set of configurations defining the 
transition-state, while
the set $a$ are configurations in the compound nucleus on one side
or the post-barrier configurations on the other side.  The interaction
connect the transition-state configurations to the rest is denoted 
$v$.  

Due to the lack of
orthogonality among the states in $\bfH$ an individual decay channel may couple 
to more than one configuration  in the transition-state channel.   
This implies that  $\bfGam$ can have 
off-diagonal matrix elements. If these off-diagonal matrix elements
are ignored,  the individual transmission factors may violate the
Eq. (\ref{bound}).
 
The matrix structure can be achieved in a generalized Fermi Golden
Rule of the form
\be
(\bfGam_j)_{kk'} = 2\pi \sum_{j\in a} \langle k |v | j\rangle \langle k' |v |
j\rangle\delta(E_j - E).
\label{gFGR}
\ee
In this work we do not attempt to compute the $\bfGam$ from Eq.
(\ref{gFGR})
from the Hamiltonian but simply assume the separable approximation
\be 
(\bfGam_j)_{k k'} = \gamma_j g_{j}(k) g_{j}(k').
\label{separable}
\ee
to parameterize it.  

\section{Examples of GOA Hamiltonians}
\label{GOA}
In these examples the transition-state channel is composed of
one or two chains of configurations with varying assumptions
about the Hamiltonian $\bfH$.  Calling the collective GCM
coordinate $q$, we take chains of $N$ regularly spaced states
spanning an interval $[q_1,q_N]$ with a mesh spacing $\Delta q = 
(q_N - q_1)/(N-1) = 1$.  To examine the dependence on the mesh 
spacing the calculations are carried out for two choices of
mesh spacing, keep the Hamiltonian and the total length 
$Q_T = (N-1) \Delta q = 3$ fixed.  

To derive the matrix elements 
in the model, we assume that the variables
in the wave function of a configuration $\Psi_{q_k}$ can be decomposed into a
continuous coordinate $q$ and a set of other coordinates
$\xi$, and the dependence on $q$ is Gaussian in the GCM
configuration,
\be
\Psi_{q_k}(\xi,q) =  \Psi'(\xi)(\pi s^2)^{-1/4}\,e^{-(q-q_k)^2/(2 s^2)}.
\label{factorization}
\ee
Here $s$ is the width of the Gaussian wave
packet.  Then the overlap matrix has elements
\be
\bfN_{jk} = 
\exp\left(-(q_j - q_k)^2/4 s^2\right).
\ee

The model Hamiltonian matrix is constructed with separate kinetic and
potential energy terms,
\be
\bfH = \bfK + \bfV.
\label{HeqKV}
\ee
For the matrix $\bfK$, we are guided by the GCM theory for a 
cluster of particles bound together by a translationally invariant
particle-particle interaction, but free of any external forces
\cite{be19c}. Here the collective variable is $x$, the
position of the center of mass of the cluster.  Under the factorization hypothesis 
(Eq. (\ref{factorization}) the
GCM Hamiltonion matrix element is 
\be
\label{K}
\bfK_{jk} = E_K  ( 1 - (q_j-q_k)^2/2 s^2) \bfN_{jkj}. 
\ee
with $E_K$ given by
\be
E_K = \frac{\hbar^2}{2Ms^2}
\ee
and $M$ denoting the mass of the particles
\cite{be19c}.  We will
treat a possible potential energy $V(q)$ in a similar 
way in Section B below.

We make the same separable  approximation for the 
for imaginary matrices $\bfGam$, centering their
wave packets at the endpoints of the chain.
The resulting parameterization from 
Eq. (\ref{separable}) has the separable function
\be
g_{j}(k)= \bfN_{k j}
\ee
and $\gamma$ as an arbitrary real parameter.
Here we assume $\gamma_1=\gamma_N\equiv \gamma$. 

The resulting Green's function to be evaluated is
the inverse of the matrix  
\be
\bfH_{jk} - i \gamma \bfN_{j1}\bfN_{k1}/2
- i \gamma \bfN_{j N}\bfN_{k N}/2 -\bfN_{jk} E.
\ee

In summary, aside from the term $\bfV$, the
model presented here has three dimensionless parameters:  $N$, the number
states in channel; $\Delta q/s$, the spacing of the states in units of the width
of the collective wave packet; and $\gamma/E_K$, the strength of the
imaginary decay width in units of the zero-point kinetic energy. The
energy scale is set by $E_K$.  The width of the Gaussian packet $s$ is
also dimensionful and sets the scale for the overlap distance
between configurations \cite{bo90}.

\subsection{A single flat channel}

\begin{figure}[tb] 
\begin{center} 
\includegraphics[width=\columnwidth]{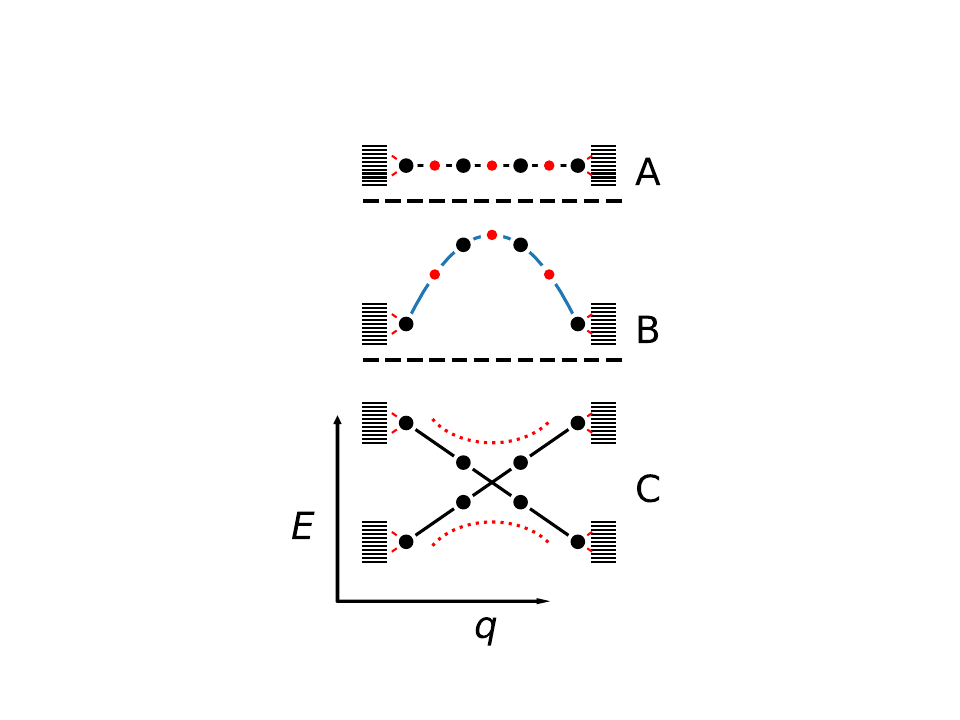} 
\caption{Relationship between states in the models described in Sections
IV.A,B, and C. The states in the 4-state and 7-state  channels are
shown as black and  black+red circles respectively.  
The real part of the Hamiltonian couples the states in
the channel or channels; the couplings to the reservoirs are parameterized
by the imaginary part of the Hamiltonian.}
\label{configs}
\end{center} 
\end{figure}

\begin{figure}[tb] 
\begin{center} 
\includegraphics[width=\columnwidth]{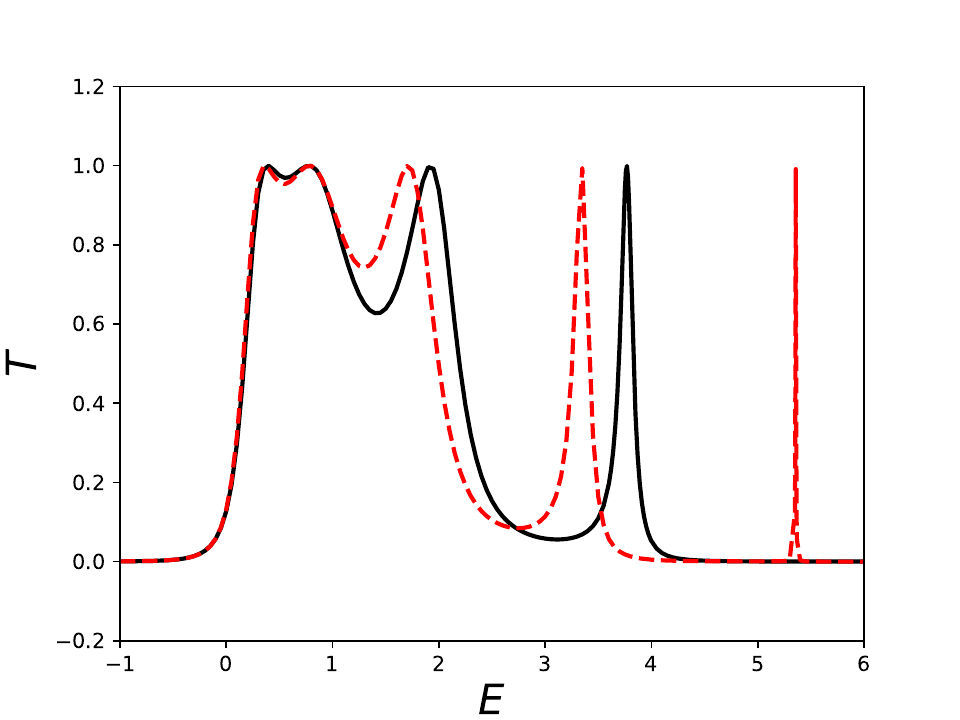} 
\caption{Transmission factor for a chain of length $Q_T =  3\sqrt{5}$
comparing GCM calculations for 4 and 7 states in the chain (solid black and
dashed red lines, respectively). Besides the peaks visible in the 
figure, there are 2 extremely narrow peaks at somewhat higher energy
in the 7-state model.  The parameters of the Hamiltonian are 
$(s,E_0,\gamma) = (1/\sqrt{5},5/4,1)$. See the Supplementary Material 
for the computer scripts used to 
calculate the data presented here and later in the Figures.} 
\label{4vs7}
\end{center} 
\end{figure}

The first model we investigate is a flat chain composed of $N=4$ configurations
with overlaps between them set to $\Delta q/s = \sqrt{5}$. This choice was
shown in Ref. \cite{be19c} to give a good compromise between accuracy and
computational effort.  The channel is depicted as ``A" in Fig.
\ref{configs}. The states indicated by black circles are the ones included in the
$N=4$ model.  We will also examine the same model with 7 states; 
the added states are shown as the red circles. For the 7-state model 
$\Delta q$ is reduced by a factor of 2 while $s$ remains fixed.
The diagonal energies of the GCM states $E_K$ are taken to be $E_K = 5/4$ and the
strengths 
of the absorption
at the ends are  $\gamma = 1$.  With these parameters the overlap between neighboring 
states is fairly small,  $\bfN_{k,k+1} = 0.28$. The resulting 
transmission factor $T(E)$ calculated
by Eq. (\ref{datta}) is 
shown in Fig. \ref{4vs7} as the black solid line.  
One sees a structure of 4 peaks,
each close to an eigenvalue of the Hill-Wheeler  Eq. (\ref{Hv=ESv}).  Physically, the peaked structure
arises from the wave reflection at the ends of the channel.  Note  
that the range of $T(E)$ satisfies $0 \leq T \leq 1$ as required by the
unitarity of the $S$ matrix.  Note also that the channel starts
conducting near
$E\approx 0$, as would be the case for a classical channel governed by a
Hamiltonian without any zero-point energy.  
The adequacy of the mesh spacing can be
assessed by shrinking it.  Decreasing it by a factor of 2, the
same interval contains 7 seven states instead of 4.  The resulting
transmission factor is shown as the dashed red line in Fig. \ref{4vs7}.  
One sees that in the low-energy region it is quite
similar to the 4-state approximation.   However, it has 3 additional peaks at 
higher energy, corresponding
to the high-energy eigenfunctions of the 7-d model.  
These peaks are much narrower than the lower ones and can be neglected in
calculating integrated transmission rates.  The same behavior
would continue with finer mesh spacings;  there would remain 4 peaks in the
energy region $[0,2]$ and the additional narrow peaks would appear at higher
and higher energies.  
The qualitative aspects of this  behavior can be easily understood.  With
a finite mesh spacing of Gaussian wave packets one can approximate plane
wave with a good fidelity for low momentum, but there is a momentum 
cutoff controlled by the mesh spacing.  In the transmission channel as
parameterized, the momentum at the injection and exit point is controlled
by the Gaussian width parameter $s$.  The momentum match to the channel
parameters suppresses the transmission to the high-momentum modes in 
the channel.  We conclude that fairly sparse meshes are adequate for
representing the overall conductivity of flat transmission channels.

As mentioned earlier, very fine mesh spacings often 
lead to numerical instabilities
in the spectroscopic applications of the GCM.  The usual fix is to
make a singular value decomposition of the overlap matrix, throwing out
eigenfunctions that have small norms.  It is instructive to see what happens
when the same procedure is applied here.
\begin{figure}[tb] 
\begin{center} 
\includegraphics[width=\columnwidth]{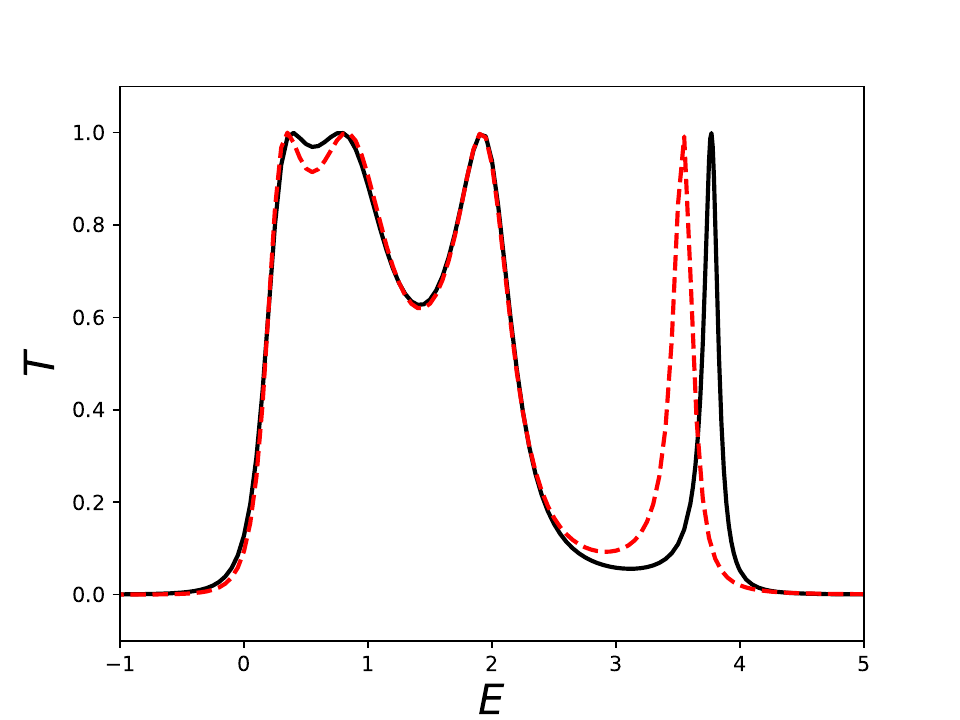} 
\caption{Transmission factor for a chain of length $q = 4$
comparing GCM calculations for 4 and 7 states in the chain (solid black and
dashed red lines, respectively).  The difference with Fig. \ref{4vs7} is that 
the 7-d space was truncated to 4 dimensions by the singular value
decomposition of the overlap matrix. 
The Hamiltonians are the same as in Fig. \ref{4vs7}.} 
\label{compare}
\end{center} 
\end{figure} 
Fig. \ref{compare} compares the $4$-state model
with the 7-state model truncated to 4 states.
That is, we diagonalize the
norm matrix in the 7-state model and project the Hamiltonian on the
basis of the 4 eigenfunctions  having the highest eigenvalues of the norm matrix. 
One sees that the resonance
positions are rather close and the widths are also very similar.
There is no obvious benefit from 
starting out with a larger space.  Since there is no need to 
truncate the space for reasons of numerical stability, this
aspect of the usual methodology can be dropped.

\begin{figure}[tb] 
\begin{center} 
\includegraphics[width=\columnwidth]{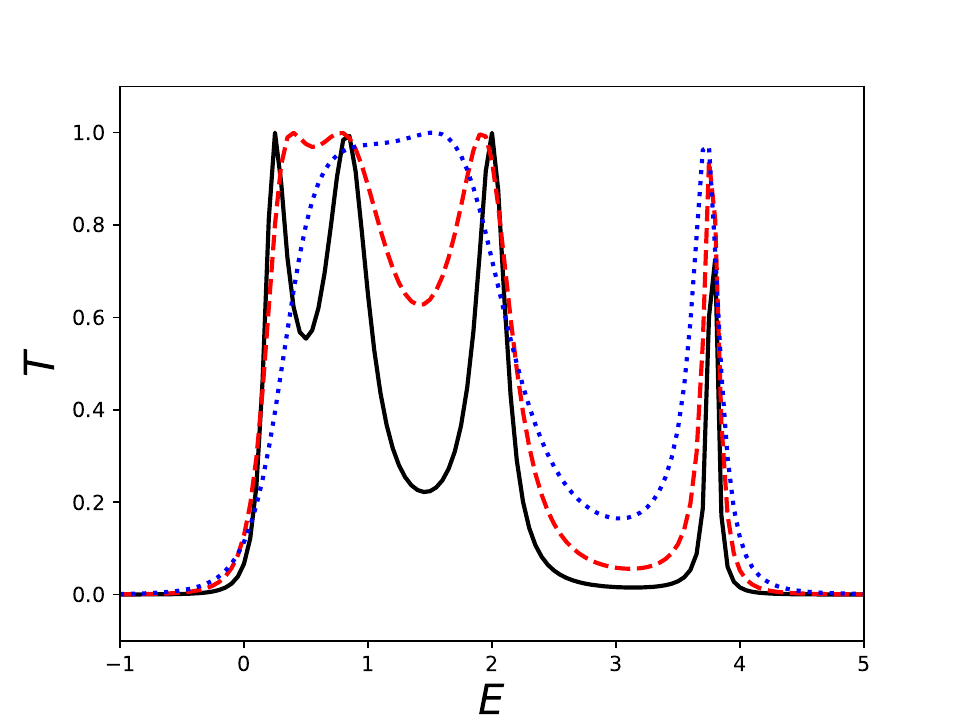} 
\caption{Transmission factors in the 4-d model for several values of
absorption strength:  $\gamma=0.5$ (solid black line);  $\gamma=1.0$ (red dashed line);
$\gamma=2.0$ (blue dotted line).  All peak tops are at $T=1$, but
may not be visible in the figure due the finite grid of energies.}
\label{Tvsg} 
\end{center} 
\end{figure}

We next examine how $T(E)$ depends on the strength of the absorption
at the ends of the channel.  Fig. \ref{Tvsg} shows $T(E)$ for a range of
absorption strengths $\gamma$.  
Obviously, for small $\gamma$ the
channel acts as a resonant cavity with sharply defined resonances and
the overall conductance is low.  For the larger $\gamma$'s 
the reflection amplitude is small and the individual peak broaden and
merge together.  

\subsection{A parabolic  channel}

In this section we extend the model to include a potential barrier.  We
take the shape of the barrier as an inverted parabola, as is
often assumed in phenomenological treatments. 
 
\begin{figure}[tb] 
\begin{center} 
\includegraphics[width=\columnwidth]{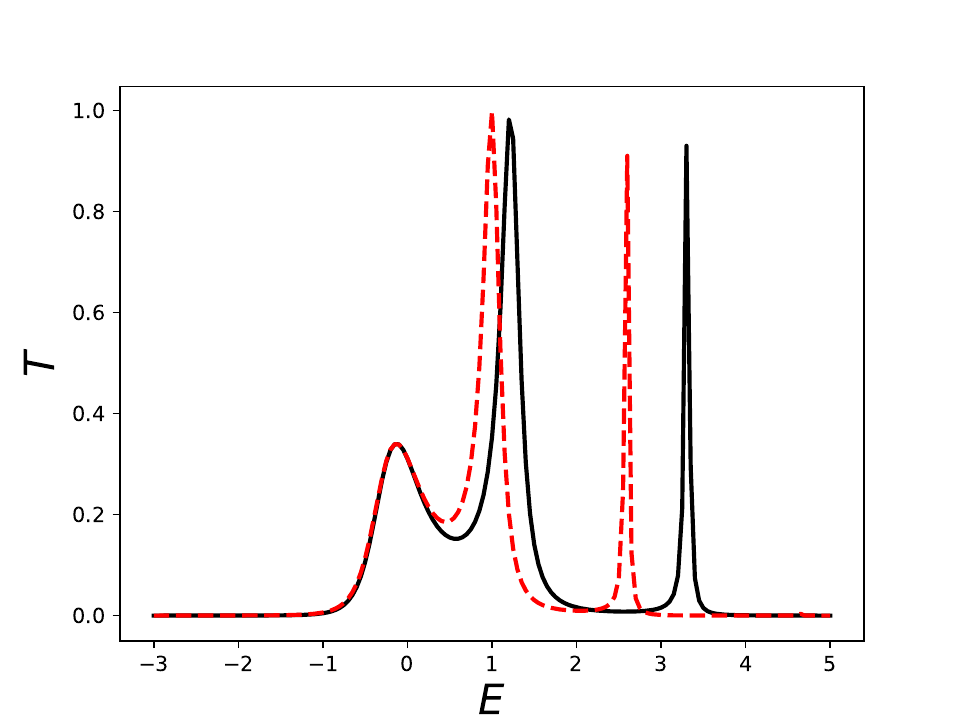} 
\caption{Transmission factor in the 4-d and 7-d models with a parabolic barrier.
Solid black line:  4x4 model; red dashed line: 7x7 model. As in the previous
figures, the narrow peaks extend up to $T=1$ in height.}
\label{4vs7V} 
\end{center} 
\end{figure}

Under the factorization Ansatz Eq.~(\ref{factorization}) the GCM matrix elements of a
potential depending only on the $q$ coordinate are given by
\begin{eqnarray}
&&\bfV_{jk} \nonumber =\\
&& \frac{1}{s\pi^{1/2}}\int^\infty_{-\infty} dq\,
V(q) e^{ - (q-q_j)^2/2s^2 - (q-q_k)^2/2s^2}. 
\end{eqnarray}
The $V(q)$ is taken as the parabolic form  
\be
V(q) = V_2 (q - q_b)^2
\ee
where $q_b$ is at the center of the barrier.  The 
resulting GCM matrix elements are
\be
\bfV_{jk}= V_2 \left[\left(\frac{q_j+q_k}{2} -
q_b\right)^2 + \frac{s^2}{2}\right]\bfN_{jk}.
\label{Vmat} 
\ee
The  matrix $\bfV$ of these elements are 
added to the Hamiltonian defined in Eq. (\ref{HeqKV}) and (\ref{K}).  
Note that the
diagonal potential matrix elements are slightly below the defining potential
due to the second term in Eq. (\ref{Vmat}).
The diagonal
energies are indicated in the channel marked ``B" in Fig. \ref{configs}.
For a numerical example we take $V_2 = -1/2$. The channel Hamiltonian
has 4 eigenenergies ranging from $-$0.4 to 3.3.  
Fig. \ref{4vs7V} shows the transmission factor as a function of energy taking
$\gamma = 1.0$.
Three peaks are visible. In terms of the channel eigenstates, the two lowest
are responsible for the broad peak at $E\approx 0$.  It appears that the
barrier suppresses the maximum conductance since the peak height much less
than one.  At  higher energies the $T$ can approach  maximum
value of one, but the coupling is weaker and the peaks are narrower.

We believe this behavior is generic for channels that follow
the topography of the potential energy surface. This is the
case when they are  constructed using an adiabatic approximation.
To see that the results are not an artifact of the GCM mesh
spacing, we also show 
the transmission factor taking a finer mesh
with 7 GCM configurations instead of 4.  One sees that the
low-energy conductance is almost the same.  At higher energies,
the narrowing of the peaks is also similar, although the 
peaks are somewhat shifted in energy.

\subsection{Two crossing channels}

To understand better the adiabatic approximation,  we consider a model in which
the adiabatic channels arise from coupling between diabatic ones.
We start with two diabatic channels that cross as depicted in graph ``C"
in Fig. \ref{configs}.  
The dashed black lines link configurations
that have large matrix elements in HF mean field theory; 
resulting chains are the diabatic paths in the dynamics.  
Adiabatic dynamics arise when one 
first diagonalizes the Hamiltonian  within the subspace
at fixed $q_k$.  These are
indicated by the curved red
dotted lines in the Figure.  The picture of adiabatic channels
peaking at the barrier top is unavoidable in transition-state
theory as implemented in Eq. (1). 

For the Hamiltonian model we add
linear potentials to generate the diabatic
paths together with a  constant interaction between configurations
states at the same positions $q_k$.
The matrix elements for a potential having a constant slope  $V_a(q) = v_a q$
are given by
\be
\bfV_{a j ;a k} = v_{a} \left(\frac{q_j+q_k}{2}-q_b\right) \bfN_{jk}\delta_{a a'}.
\ee
The label $a$ applies to the upward-sloping diabatic channel; the
downward-sloping one will be label $b$.
The other term to be added to the Hamiltonian is the coupling $h_c$ between
the configurations  of the two diabatic channels. We take the coupling as 
a constant independent of $q_k$.  Again invoking factorization hypothesis,
the nonzero matrix elements are
\be
\bfH_{aj,b k} = h_c N_{jk}.
\ee 
For the numerical example, we take $v_a=0.5$, $v_b=-0.5$ and
$h_c = 0.8$. 

As depicted in Fig. \ref{configs} there are now 4 decay matrices to be added to the 
Hamiltonian.  We assume that all final states are orthogonal to each other, so we
can apply the transmission formula Eq. (\ref{datta}) with an incoherent sum over all four
combinations $(a1,b1)\rightarrow (aN,bN)$.  

\begin{figure}[tb] 
\begin{center} 
\includegraphics[width=\columnwidth]{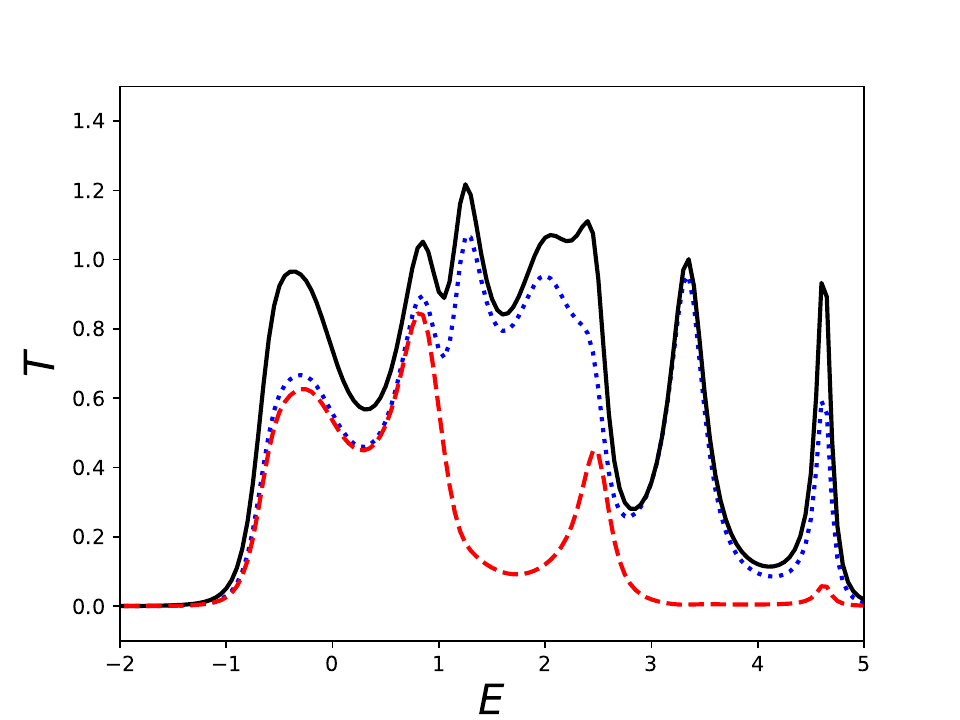} 
\caption{Transmission factor for the Hamiltonian ``C$_8$" depicted in Fig.
\ref{configs}.
Solid black line: all contributions to $T$;  red line: lowest adiabatic
contribution; blue line: both adiabatic contributions.  Note that the
total transmission coefficient can be larger than one in the cases where
two channels contribute.} 
\label{crossing2} 
\end{center} 
\end{figure}

In the adiabatic approximation only the transmission factor from the two
lowest states at the ends are included, 
\be
T_{\rm adiabatic} \approx T_{a1,bN}.
\ee
It is shown as the red dashed line in Fig. \ref{crossing2}.  
The dotted blue line shows the combined transmission factor that includes
including the upper adiabatic channel as well,
\be
T'_{\rm adiabatic} \approx T_{a1,bN} + T_{b1,aN}.
\ee
These are to be compared to the full transmission factor (solid black line) including all
contributions,
\be
T = T_{a1,bN} + T_{a1,aN} + T_{b1,aN} + T_{b1,bN}. 
\ee
One sees that the adiabatic approximation works well overall when both
channels are included.  The second channel adds hardly anything in the 
region where
the lower channel is open, but fills out the higher region $ 1.0 < E < 2.5$. 
Another interesting finding, not very visible in the figure, is that 
the adiabatic approximation significantly 
underpredicts the transmission factor at the lowest energies.  This
inadequacy of the approximation was noted earlier in Refs.
\cite{ha20,gi14}.

We have also calculated $T$ without any coupling between the diabatic
channels. As expected, that treatment seriously underpredicts the 
transmission coefficient.

\section{General conclusions}.

A few tentative conclusions may be drawn from the simple models 
presented here.  First of all,  one does not need fine
collective-coordinate meshes in the GCM configuration space.  A
mesh spacing giving overlaps of 0.3 between a configuration and its
diabatic neighbor seems adequate;  smaller mesh spacings will 
produce narrow resonances at high energy, but the coarse properties of the
channel will remain the same.  The second conclusion is that momentum
matching is an important consideration in the channel coupling to
the reservoir states.  It produces an effective energy cutoff in the
conductance of the channel.  The energy scale for this effect is
given by the zero-point energy of the collective coordinate in
the mean-field wave function.  To give a sense of that, we present
in Table I some characteristics of the transmission function $T(E)$
for the models discussed in the previous section.  
The first characteristic is the integrated
transmission factor.  This is reported in the Table in units of $E_K$,
\be
I_T = \int_{-\infty}^\infty d E \, T(E)/E_K.
\ee
Comparing the 4-state models
with the 7-state models A and B, one sees less than a 10\% change
in the integrated transmission.

A second finding is that the transmission is strong only in a limited energy interval.
To examine this point in a quantitative way, we have computed the $T$-weighted
average energy 
\be
\langle E \rangle = \left.\int d E \, E \,T(E)\right/\int d E \, T(E)
\ee
and its standard deviation, $\sigma(E) = \langle E^2\rangle - \langle E
\rangle^2$.  These quantities are shown in the last two columns of the
table.
\begin{table}[htb] 
\begin{center} 
\begin{tabular}{|c|ccc|} 
\hline 
Model  &   $I_T$  & $\langle E\rangle/E_K$ & $\sigma(E/E_K)$     \\
\hline 
A$_4$ &  1.69 &  1.17 & 0.81 \\ %
A$_7$ &  1.65 &  1.11 & 0.77 \\       
B$_4$ &  0.61   & 0.80 & 0.84 \\
B$_7$ &  0.56   & 0.60 & 0.67 \\
C$_8$ &  3.12   & 1.28 & 1.15 \\
C$^a_8$ & 2.54 & 1.31  & 1.09 \\
\hline 
\end{tabular} 
\caption{Integrated channel properties of the models discussed in
the text.  The model labels refer to the subsection in Sect. IV where
they were discussed. The subscript refers to the dimension of the GCM
space.  The average energy $\langle E \rangle$ is computed with respect
to the Hamiltonian at the midpoint of the $q$ interval omitting the
zero-point energy $E_K$. 
In case C, the energy is the adiabatic one
computed by diagonalizing the 2x2 matrix mixing the two
GCM states. The row marked C$^a_8$ ignores the coupling between
the adiabatic channels.}
\label{exact} 
\end{center} 
\end{table} 

Both quantities exhibited in the Table support our conclusion that
one can safely use coarse meshes
to define the channels. The averages change by 25\% or less in comparing
the 4-state and 7-state models.  The average energy is comparable
to the zero-point energy in the flat channel. In the parabolic channel the
average energy is somewhat lower due barrier effect;  in this case the finer mesh has  
a significant affect.  However, it may be seen from the $\sigma(E/EK)$ 
column that the spread of energies is about
the same.  

Model $C_8$ simulating adiabatic transmission by two interacting diabatic channels
has about
twice the integrated transmission strength as model A,
which is hardly surprising.  Note however that the coupling between
the two diabatic channels is significant:  model $C^a_8$ ignores
the coupling and its $I_T$ is smaller by 20\%.  In present fission
theory such couplings are neglected, and this finding confirms that
approximation in phenomological models.

\end{document}